\documentclass[5p]{elsarticle}

\usepackage{graphicx, subfigure}
\usepackage{amsmath}
\usepackage{amsfonts}
\usepackage{amssymb}
\usepackage{slashed}
\usepackage{hyperref}
\usepackage{amsbsy}
\usepackage{natbib}
\usepackage{balance}

\newcommand{\be}{\begin{equation}}
\newcommand{\ee}{\end{equation}}
\newcommand{\ba}{\begin{eqnarray}}
\newcommand{\ea}{\end{eqnarray}}

\journal{Elsevier}

\begin{document}

\begin{frontmatter}

\title{Conformal or Confining}

\author[rug]{T. ~Nunes da Silva}
\ead{t.j.nunes@rug.nl}

\author[rug]{E. ~Pallante\corref{cor1}}
\ead{e.pallante@rug.nl}

\author[rug]{L. ~Roebroek}
\ead{l.l.robroek@student.rug.nl}

\cortext[cor1]{Corresponding author}

\address[rug]{Van Swinderen Institute for Particle Physics and Gravity, Nijenborgh 4, 9747 AG, Groningen, The Netherlands}

\begin{abstract}
We present a lattice study of the phase transitions at zero and nonzero temperature for the $SU(3)$ gauge theory with a varying number of flavours $N_f$ in the fundamental representation of the gauge group. We show that all results are consistent with a lower edge of the conformal window between $N_f\!=\!8$ and $N_f\!=\!6$. A lower edge in this interval is in remarkable agreement with perturbation theory and recent large-$N$ arguments.  
\end{abstract}


\end{frontmatter}

\section{Introduction}
Physicists are familiar with two paradigms for conformal symmetry loss in four-dimensional interacting quantum field theories:  Quantum Electrodynamics (QED), whi\-ch is infrared free, and Quantum Chromodynamics (QCD), which loses conformal symmetry in a highly non trivial way, leading to two manifestations of one single breaking phenomenon: asymptotic freedom and confinement \footnote{Confinement in this context is the existence of a mass-gap, i.e., a non-zero glueball mass in quenched QCD.}.

A richer dynamics can be realised in the presence of non-trivial, i.e., interacting, ultraviolet or infrared fixed points of the quantum theory, and may turn out to play a role in unifying the standard model of particle physics with gravity. The possibility that such a scenario is realised somewhere between the electroweak symmetry breaking scale and the Planck scale has recently attained strong theoretical and experimental appeal. The search for re\-norm\-ali\-zation-group fixed points in candidate theories for particle physics and cosmology also shares a wide range of tools and aspects  with the study of quantum critical phenomena in condensed matter systems, in three as well as lower spatial dimensions. 

The simplest step beyond the QCD paradigm is the so-called conformal window: a family of zero-temperature theories in the phase diagram of non-Abelian gauge theories ranging over a number of massless flavours $N_f^c\!<\!N_f\!<\!N_f^{UV}$ in a given representation of the gauge group. The critical number of flavours $N_f^c$ marks its lower edge, and $N_f^{UV}$ the upper edge where ultraviolet freedom is lost. The beta-functions $\beta(g)$ of theories inside the conformal window have a zero at some coupling $g\! =\! g^*$, where the theory has a non-trivial infrared fixed point (IRFP) \cite{Caswell:1974gg, Banks:1981nn} and it is conformal; they are negative for $0\!<\!g\!<\!g^*$, and zero at $g\!=\!0$ (ultraviolet freedom). Above the conformal window, $\beta (g)\!>\!0$ for $g\!\gtrsim\!0$ implies infrared freedom opening up the interesting possibility of asymptotic safety, i.e., a non-trivial ultraviolet fixed point (UVFP) \cite{Litim:2014uca}. 

Many questions related to non-trivial fixed points and  the emergence of the conformal window\,---\,its location in the parameter space and accompanying signatures\,---\,are genuinely nonperturbative, and require non\-per\-tur\-ba\-tive strat\-e\-gies. Gauge/gravity duality \cite{Aharony:1999ti} can come to rescue whenever the gauge theory is conformal, as in the conformal window, or near-conformal and still deconfined, as in the high-temperature quark-gluon-plasma. This seems to be not true in the confining and asymptotically free phase of QCD \cite{Bochicchio:2016euo} that precedes the conformal window; the recently proposed solution \cite{Bochicchio:2013eda} for the scalar glueball correlator in the 't Hooft limit of large-$N$ QCD tells that its momentum dependent logarithms cannot be reproduced by AdS/CFT realisations so far.

In this letter we constrain the lower edge of the conformal window for the $SU(3)$ gauge theory with $N_f$ flavours in the fundamental representation (i.e., many-flavour massless QCD) by means of a lattice study, a genuinely nonperturbative approach, tailored to the physics problem at hand. Knowing the location of the lower edge is one way to establish how far large-$N$ predictions, or perturbation theory to a given loop order are from the complete theory. It is also essential in order to correctly identify the properties of many phenomenological models for particle physics with a composite spectrum. 
\section{The lower edge of the conformal window }
\label{sec:results}
\begin{figure}[t]
\centering
\includegraphics[width=.95\columnwidth]{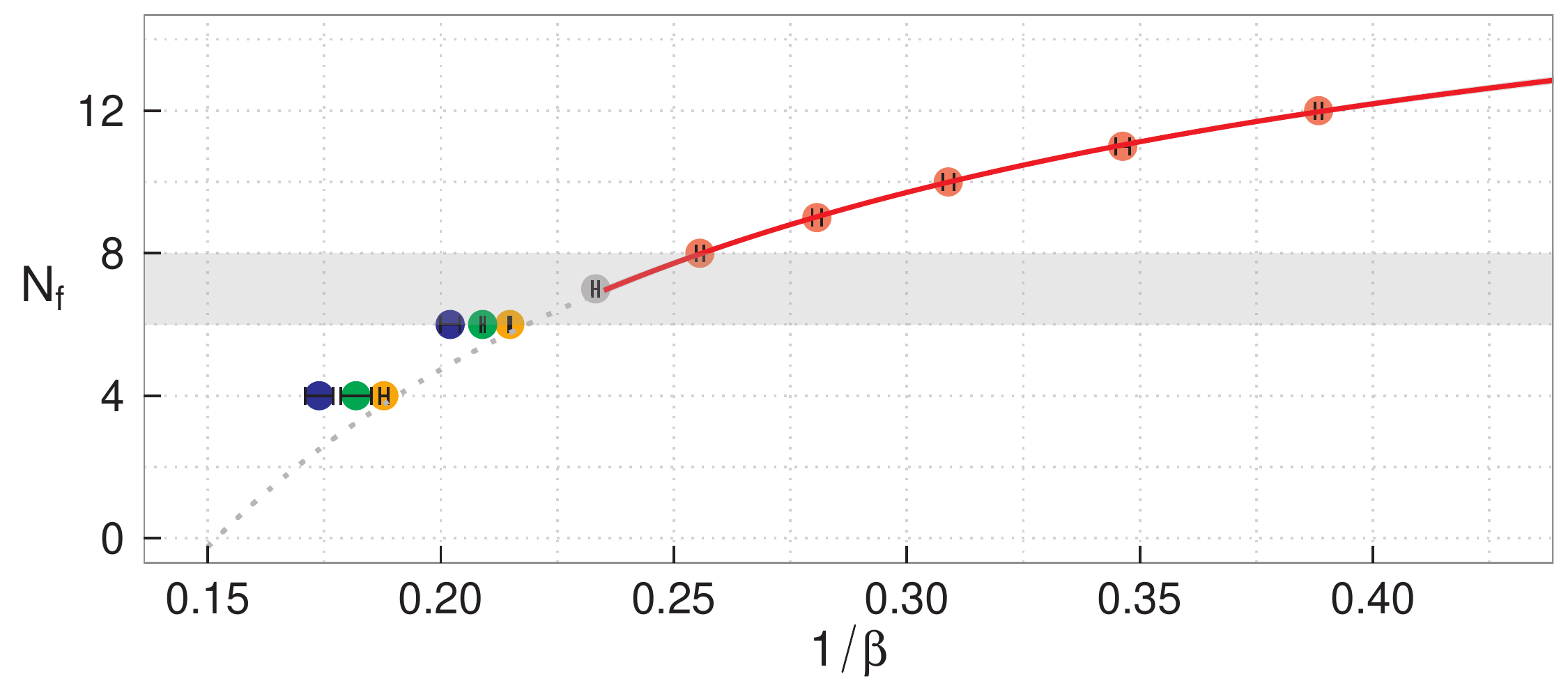}
\caption{\label{fig:summary} Lattice phase diagram $N_f-1/\beta$, $\beta\!=\!10/g_L^2$, $g_L$ the lattice coupling. 
Data for $N_f\!=\!8,\ldots ,12$ (red circles) are midpoints of sharp zero-temperature crossovers in Fig. \ref{fig:bulk}. Best fit to data (solid red) identifies the linear in $\beta$ bulk line, extrapolated to $N_f\!<\!8$ (dotted grey). $N_f\!=\!7$ crossover in Fig. \ref{fig:bulk} falls on the line (grey circle). Bound on $N_f^c$ from this work is sketched (grey band). 
$N_f\!=\!4,\, 6$ thermal transitions occur away from the bulk line, with volumes $24^3\times N_t$, $N_t\!=\!6$ (orange), $8$ (green), and $12$ (blue), right to left.     }
\end{figure}
Figure  \ref{fig:summary} summarises the results of this work and explains its strategy: theories with $N_f\!<\!N_f^c$ have spontaneously broken chiral symmetry below a critical temperature $T_c\!>\!0$, above which it is restored. Inside the conformal window, theories have exact chiral symmetry and are deconfined at all temperatures. Hence, one way to determine $N_f^c$ is to follow the chiral symmetry breaking pattern, with varying $N_f$ and temperature, for the theory regularised on a Euclidean spacetime lattice. For $N_f\!<\!N_f^c$, a thermal chiral symmetry restoring phase transition occurs at some $T_c\!=\!1/(a(g_L^c) N_t)\!>\!0$, with $a(g_L^c)$ the lattice spacing at the critical lattice bare coupling $g_L^c$ on a lattice volume $N_l^3\!\times\! N_t$, with temporal extent $N_t\!\ll\! N_l$. For $N_f\!\geqslant\! N_f^c$, the lattice theory exhibits exact chiral symmetry at all temperatures, i.e., any $N_t$, and for all values of $g_L$ in the interval $0\!\leqslant\!g_L\!\leqslant\!g_L^*$, with $g_L^*$ a sufficiently large coupling where chiral symmetry will eventually be broken. This is the bulk phase transition.
 
The red solid line in Fig. \ref{fig:summary} is the line of such transitions for varying $N_f$. In any given renormalisation scheme, it manifests the fact that fermion screening is increasingly effective for increasing $N_f$ inside the conformal window; this $N_f$ dependence is a leading order effect separating two phases with different underlying symmetries, different in nature from any lattice artefact that could occur for $N_f\!<\!N_f^c$ inside a single chirally broken phase on coarse lattices. At finite lattice spacing $a$, the bulk line can be seen as the $N_t\!=\!1/(aT)\!\to\!\infty$ limit of an $N_t$-finite family of chiral phase transitions that exists for all $N_f\!>\!0$. The line therefore stops at $N_f^c$, because no $T\!=\!0$ chiral phase transition occurs for $N_f\!<\!N_f^c$.
\begin{figure}[tbp]
\centering
\includegraphics[width=.95\columnwidth]{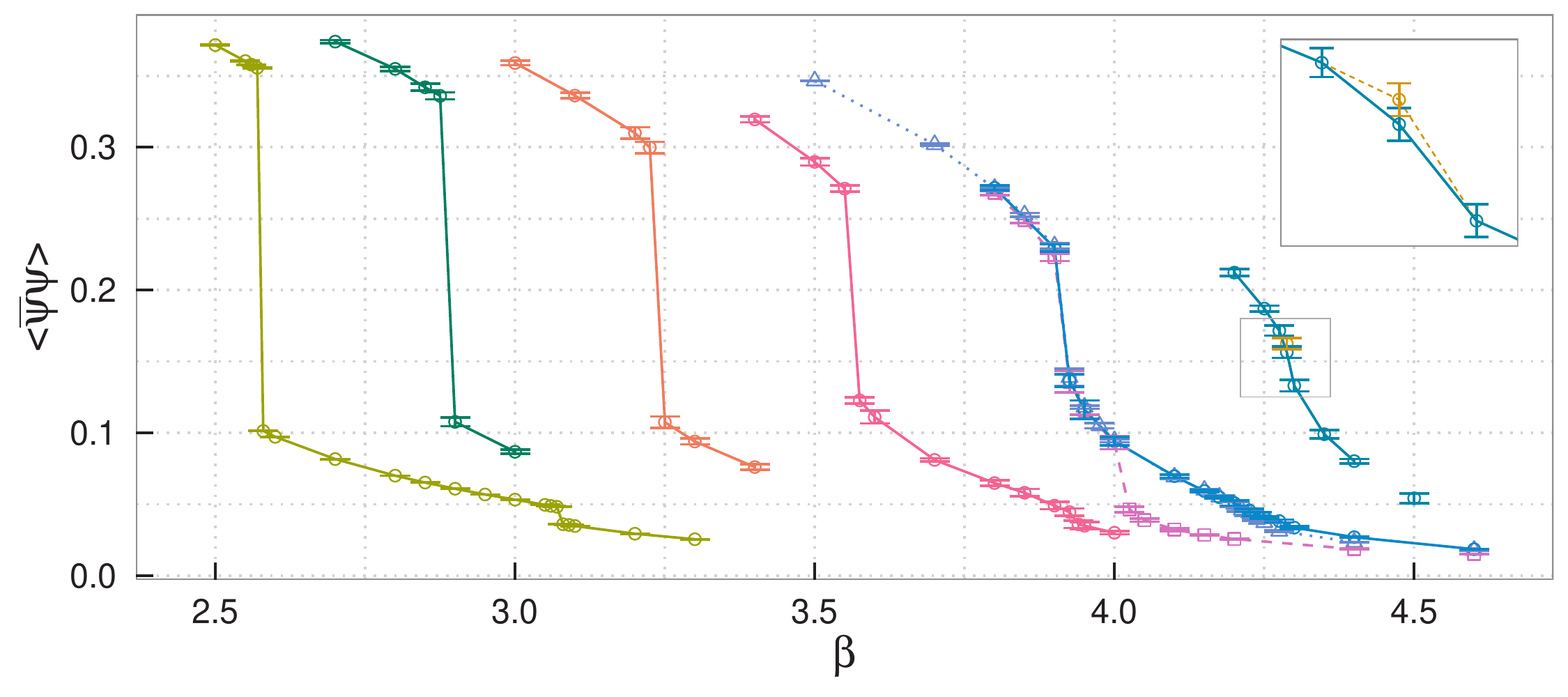}
\caption{\label{fig:bulk} Sharp crossovers of the chiral condensate (lattice units) for $N_f\!=\!12$ \cite{Deuzeman:2012ee} to $7$ (left to right). 
 $N_f\!=\!11,\,$$ 10,\,$$9$ at $12^3\!\times\!24$,  $N_f\!=\!8$ at $16^3\!\times\!32$ (circles, solid line), $24^3\!\times\!12$ (triangles, dotted line), $24^3\!\times\!6$ (squares, dashed line), $N_f\!=\!7$ at $16^3\!\times\!32$; all with bare lattice mass $am\!=\!0.01$.  
Inset shows data for two branches of possible hysteresis loop for $N_f\!=\!7$. 
Edge of exotic phase displayed for $N_f\!=\!12,\,9$, and $N_f\!=\!8$.
}
\end{figure}

With this paradigm, we have studied the $N_t$ dependence of chiral phase transitions for $4\!\leqslant\!N_f\!\leqslant\!12$ in QCD.  The setup of the simulations is the one of \cite{Lombardo:2014pda}. Throughout, we have not observed an anomalous behaviour that could hint at consequences of the fourth root of the fermion determinant for staggered lattice fermions. We have studied the chiral condensate $\langle\bar{\psi}\psi\rangle$, order parameter of spontaneous chiral symmetry breaking, the disconnected chiral susceptibility $\chi_{\text{disc}}$, and the con\-nec\-ted chiral cumulant $R_\pi$=$\chi_{\text{conn}}$ /$\chi_\pi$=$(\partial \langle \bar{\psi}\psi\rangle$/$\partial m_{\text{valence}})$/$(\langle \bar{\psi}\psi\rangle / m)$. This quantity is a powerful probe of chiral symmetry: in the chirally broken phase, the pseudoscalar lowest-lying state is a Goldstone boson and its vanishing mass in the chiral limit, together with the non-degeneracy of chiral partners, guarantees that $R_\pi\!\to\!0$ in the chiral limit; in the chirally restored phase, the degeneracy of the scalar and pseudoscalar chiral partners implies that $R_\pi\!\to\!1$ in the chiral limit. We have also studied the Polyakov loop $L$ at finite temperature. Despite it not being a true order parameter in the presence of fundamental fermions, it is observed to typically retain the correct features related to a deconfined or a confined phase. We used volumes with $N_t\!<\!N_l$ and aspect ratio $N_l/N_t\!\geqslant\!2$,  as well as zero temperature volumes with $N_t\!=\!2N_l$, at a fixed bare fermion mass $am\!=\!0.01$. 

Remarkably, the sharp crossovers in Fig. \ref{fig:bulk} for $N_f\!=\!12$ to $N_f\!=\!8$ are $N_t$ independent. We do observe a reduced crossover for $N_f\!=\!7$, an almost closing gap, and signs of hysteresis in the interval $\beta\!=\![4.275, 4.3]$ right on the bulk line of Fig. \ref{fig:summary}. We defer the fate of $N_f\!=\!7$ for a future study. We have also extrapolated the bulk line to $N_f\!=\!16$ obtaining  $\beta_c\!\sim\!1.2$, consistently with previous studies \cite{Deuzeman:2012ee}  carried out with different lattice actions and heavier fermions. 

A genuinely different behaviour is observed for the $N_f\!=\!4,\, 6$ theories, whose markedly $N_t$-dependent chiral symmetry breaking crossovers for varying $N_t\!<\!N_l$ are reported in Fig. \ref{fig:summary}. Already at $N_t\!=\!6$, the crossover occurs at a coupling weaker than the one predicted by the linear in $\beta$ extrapolation of the bulk line. It moves to weaker coupling for increasing $N_t$, consistently with a thermal nature of the transition. At the same time, no zero-temperature ($N_t\!\geqslant\!N_l$) phase transition or analytic crossover is observed to occur along the extrapolated bulk line, nor at weaker coupling. 

These results provide evidence that the lower edge of the conformal window lies between $N_f\!=\!8$ and $N_f\!=\!6$, the main result of this work. In what follows, we further scrutinise the $N_f\!=\!8$ and $N_f\!=\!6$ theories, discuss implications and future directions.  
\section{\label{sec:nf8} The $N_f\!=\!8$ theory and the exotic phase  }
\begin{figure}[tbp]
\centering
\includegraphics[width=.95\columnwidth]{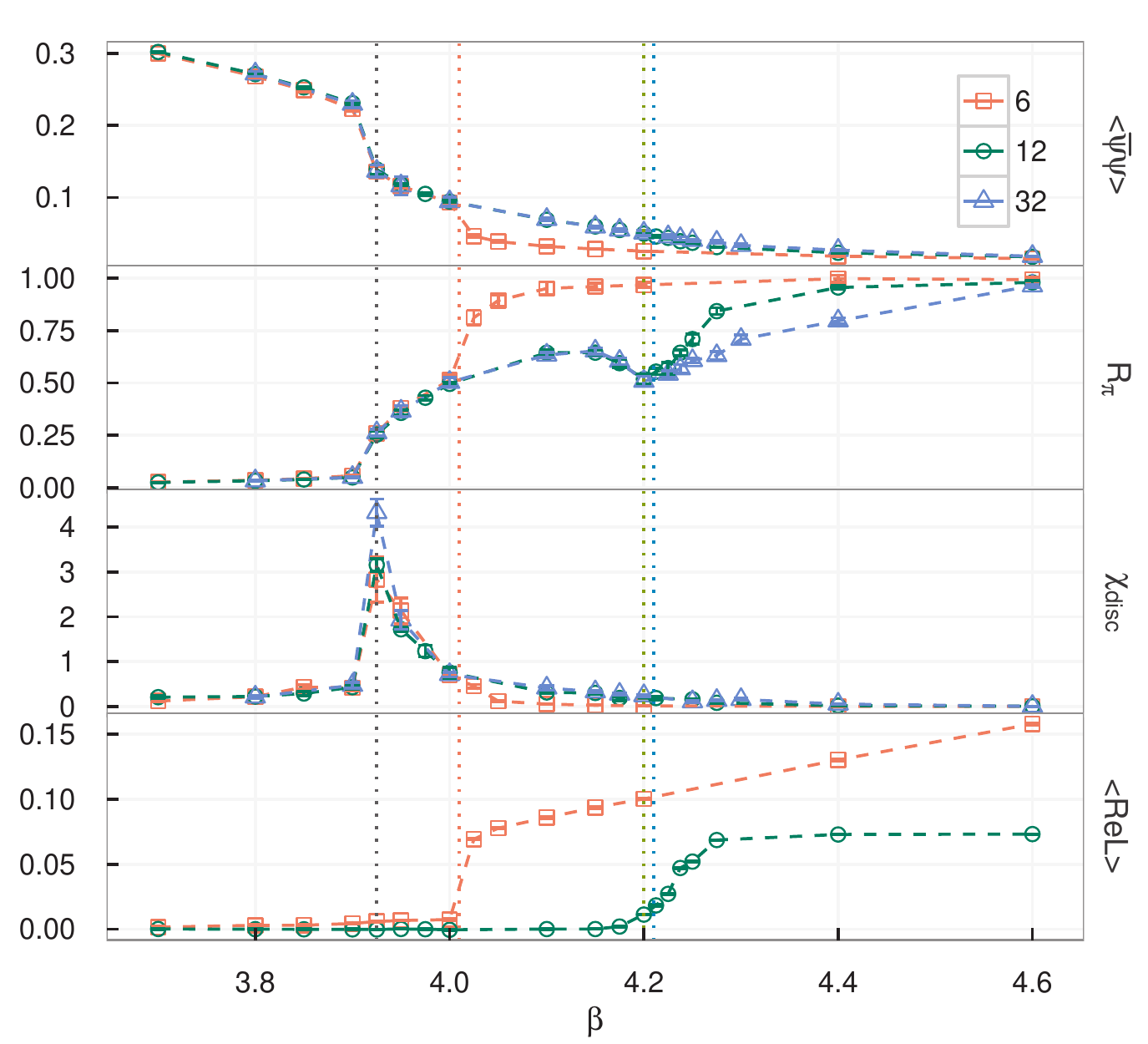}
\caption{\label{fig:nf8_4obs} 
 $N_f\!=\!8$ observables (lattice units), top to bottom: $\langle\bar{\psi}\psi\rangle$, connected $R_\pi$, $\chi_{\text{disc}}$, and $\langle Re{L}\rangle$, $24^3\!\times\! N_t$, $N_t\!=\!6$ (red squares), $N_t\!=\!12$ (green circles), and $16^3\!\times\!32$ (blue triangles), mass $am\!=\!0.01$. $\langle Re{L}\rangle$, $N_t\!=\!12$ (HYP-smeared) rescaled for better visibility.
Vertical lines indicate the location of the bulk transition as in Fig. \ref{fig:summary}, and the edge of the exotic phase for $N_t\!=\!6,\,12,\,32$, left to right [We have obtained identical results with volumes $16^3\!\times\!32$ and $12^3\!\times\!24$, which suggests that finite volume effects are under control for $N_f\!=\!8$, as well as $N_f\!>\!8$]. 
 }
\end{figure}

Figure \ref{fig:nf8_4obs} shows strikingly the $N_t$ independence of the strong coupling crossovers of $\langle\bar{\psi}\psi\rangle$ and their perfect overlap with those of $R_\pi$, and with the $N_t$-independent peaks of  $\chi_{\text{disc}}$. These signals are consistent with exact chiral symmetry for the $N_f\!=\!8$ theory at all temperatures, and the occurrence of a chiral symmetry breaking bulk transition at sufficiently strong coupling, as reported in Fig. \ref{fig:summary}. For $N_f\!=\!12$, an extensive analysis  for varying  volumes and fermion masses has suggested a first order nature of the  transition also away from the chiral limit, for a bare fermion mass $am\!=\!0.01$ \cite{Deuzeman:2012ee, daSilva:2012wg}.

Towards weaker coupling, Fig. \ref{fig:bulk} and Fig. \ref{fig:nf8_4obs} show a sequence of $N_t$-dependent reduced crossovers of the chiral order parameter, coinciding with a dip of the connected $R_\pi$ for all $N_t\!\lesseqgtr\!N_l$, and accompanied by a sharp crossover in the finite temperature ($N_t\!<\!N_l$) Polyakov loop, which remains zero at stronger coupling. We note that this second crossover in $\langle\bar{\psi}\psi\rangle$ is clearly visible for $N_t\!=\!6$, whilst its signal is suppressed at larger $N_t\!=\!12,\!32$, due to fermion mass effects at weaker coupling for fixed $am$. Two facts are important. Firstly, signals survive the zero temperature limit, and their $N_t$ dependence ceases for $N_t\!\geqslant\!12$; this excludes their thermal nature. Secondly, the absence of corresponding $\chi_{\text{disc}}$ peaks is consistent with restored chiral symmetry at all temperatures, and coupling smaller than the bulk coupling. These signatures are recognised to identify the weak coupling edge of the so-called exotic phase, a genuine lattice artefact extensively studied \cite{Deuzeman:2012ee, Cheng:2011ic} in the $N_f\!=\!12$ theory, where it is  observed to precede the bulk transition when  improvement of the lattice fermion action is used \cite{Deuzeman:2012ee}. They also suggest that the transition interpreted as a thermal transition for $N_f\!=\!8$ in \cite{Deuzeman:2008sc, Miura:2012zqa}\,---\,with a different improvement of the fermion action\,---\,is instead the edge of the exotic phase, and that both $N_f\!=\!8$ and $N_f\!=\!12$ are inside the conformal window.

That the $N_f\!=\!12$ theory is inside the conformal window and has exact chiral symmetry at weak coupling has been corroborated by the vanishing of the infinite volume chiral condensate in the chiral limit \cite{Deuzeman:2009mh} and a thorough universal scaling study of the spectrum of the theory \cite{Lombardo:2014pda}, including all lattice results available at the time \cite{Aoki:2012eq, Cheng:2013xha}. The study of $N_f\!=\!12$ in \cite{Deuzeman:2012ee}, see also \cite{Deuzeman:2011pa}, has also provided consistent evidence that the exotic phase is a chirally symmetric phase, where the U(1) axial symmetry is also effectively restored. 

One might wonder whether the improvement of the lattice fermion action could have spoiled the thermal behaviour of a theory below the conformal window, and the exotic phase as we observe it could have arisen inside a chirally broken phase. The answer is no, according to the study of $N_f\!=\!12$ with and without improvement \cite{Deuzeman:2012ee}; both cases indicate that chiral symmetry is not spontaneously broken and no thermal phase transition occurs. Furthermore, the lattice spacing $a_*$ at which the exotic phase would emerge, and the lattice spacing $a_c$ at which a thermal phase transition would occur, carry an a priori different $N_f$ and $N_t$ dependence. It would be highly accidental to have $a_*\!<\!a_c$ for any studied $N_t$ and all $N_f\!\geqslant\!8$, in order to remove traces of a thermal transition, whilst preserving the shape of $N_t$-fixed lines for all $N_f\!\leqslant\!6$. 

The exotic phase observed here has its seeds in the conformal window, which genuinely differs from QCD. It appears inside the chirally symmetric Coulomb phase of the lattice regularised theory (see Figure 2 of \cite{Lombardo:2014pda}), 
 where no continuum limit can be taken and improvement\,---\,used to reduce lattice artefacts at weak coupling\,---\, is not even approximately justified, because the theory is no longer ultraviolet free. It has a truly exotic nature: it is chirally symmetric but behaves as confining, oppositely to, e.g., finite temperature QCD with adjoint fermions, where a deconfined but chirally exact phase is allowed and indeed observed. We add that this and similarly originated exotic phases may acquire physical relevance in discrete ANNNI models and systems such as graphene in varying dimensions. A complete microscopic understanding of the exotic phase is, however, still lacking. 
\section{\label{sec:nf6} The $N_f\!=\!6$ theory and asymptotic scaling}
\begin{figure}[tbp]
\centering
\includegraphics[width=.95\columnwidth]{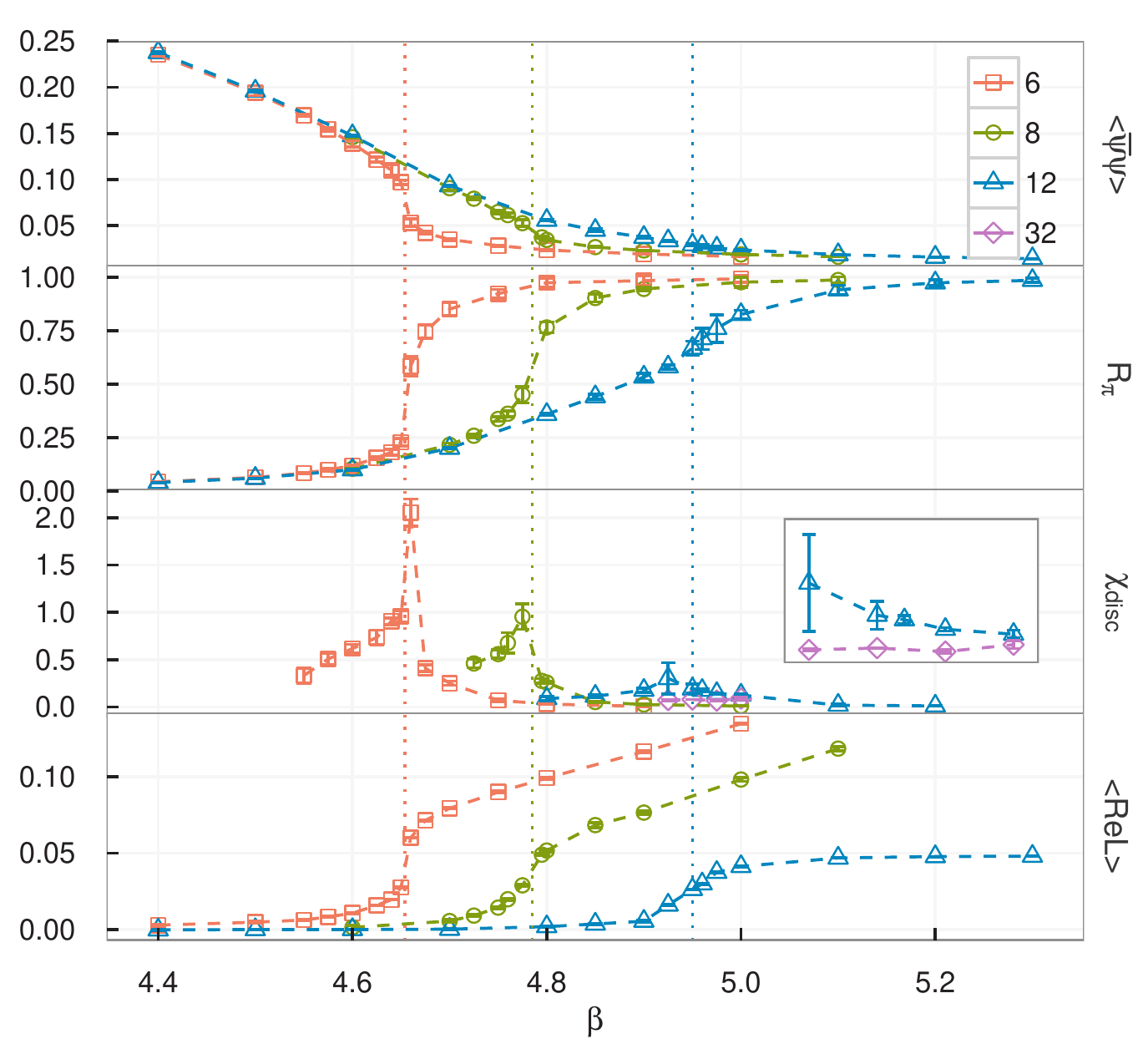}
\caption{\label{fig:nf6_4obs} 
$N_f\!=\!6$ observables (lattice units), top to bottom: $\langle\bar{\psi}\psi\rangle$, connected $R_\pi$, $\chi_{\text{disc}}$, and $\langle Re{L}\rangle$, $24^3\!\times\!N_t$, $N_t\!=\!6$ (red squares), $N_t\!=\!8$ (green circles), $N_t\!=\!12$ (blue triangles) and $16^3\!\times\!32$ (parse diamonds), mass $am\!=\!0.01$. 
$\langle Re{L}\rangle$, $N_t\!=\!8$ and $N_t\!=\!12$ (HYP-smeared) rescaled for better visibility.
 Vertical lines indicate the locations of the thermal crossovers as in Fig. \ref{fig:summary}, $N_t\!=\!6,\,8,\,12$, left to right.  
 Inset for $\chi_{\text{disc}}$: enlarged $N_t\!=\!12$ peak versus zero temperature flat distribution.}
\end{figure}

A study analogous to $N_f\!=\!8$ reveals a different symmetry pattern for the $N_f\!=\!6$ and $N_f\!=\!4$ theories. Figure \ref{fig:nf6_4obs}
shows the $N_t$-dependent signal of the chiral observables and the Polyakov loop for varying $N_t\!<\!N_l$ for $N_f\!=\!6$, and no exotic phase. 
All crossovers overlap with the $\chi_{\text{disc}}$ peaks. For a comparison, the $N_t\!=\!6$ signals for $N_f\!=\!6$ in Fig. \ref{fig:nf6_4obs} and $N_f\!=\!8$ in Fig. \ref{fig:nf8_4obs} are equally sharp, but with genuinely different characteristics. 

\begin{figure}[tbp]
\centering
\includegraphics[width=.95\columnwidth]{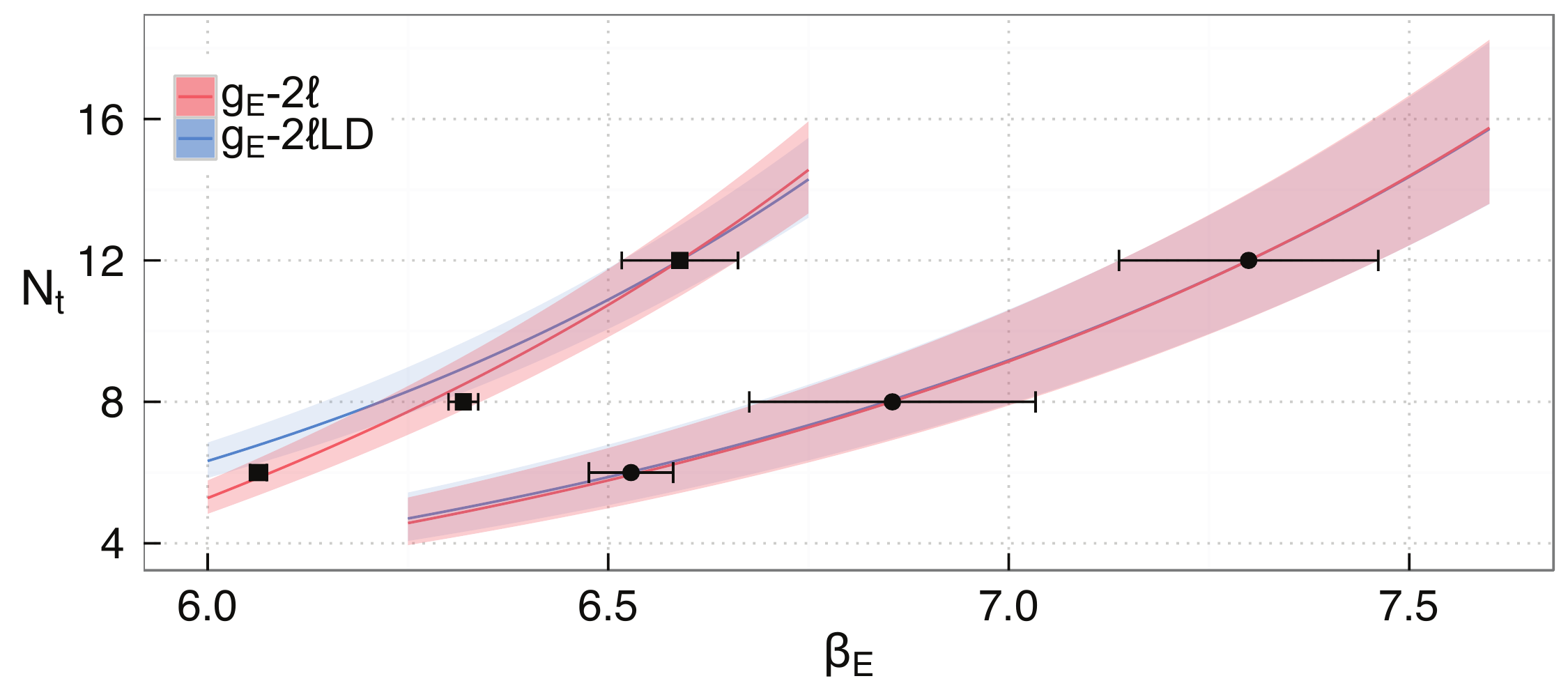}
\caption{\label{fig:nf6_asymptscaling} Lattice data with $N_t\!=\!6,\,8,\,12$, $N_f\!=\!6$ (left) and $N_f\!=\!4$ (right) as a function of the $E$-scheme improved coupling
 $\beta_E\!=\!10/g_E^2$, compared to the 2-loop scaling curve ($g_E$-$2l$) in equation (\ref{eq:scaling}), $\mbox{const}\!=\!12*R(g_E^c(N_t\!=\!12))$, and the lattice-distorted curve ($g_E$-$2l$LD) in equation (\ref{eq:scaling_LD}), $\mbox{const}\!=\!12*R_{LD}(g_E^c(N_t\!=\!12))$, $h\!=\!0.006$ ($N_f\!=\!6$) and $h\!=\!0.0005$ ($N_f\!=\!4$). Error bands are scaling curves for the extremes of the error on $g_E^c(N_t\!=\!12)$.}
\end{figure}

The thermal nature of the $N_f\!=\!6$ and $N_f\!=\!4$ crossovers is corroborated by the two-loop asymptotic scaling study in Fig. \ref{fig:nf6_asymptscaling}, similar in spirit to \cite{Miura:2012zqa}. In particular, we compare the lattice data to the two-loop scaling curve ($g_E$-$2l$)
\begin{equation}
\label{eq:scaling}
R(g_E) N_t=\left (\frac{T_c}{\Lambda_E}\right)^{-1} = \mbox{const}\, ,
\end{equation}
with $R(g_E)$(=$a\Lambda_E$) the two-loop asymptotic scaling function and the E-scheme improved coupling \cite{Parisi1981}
\begin{equation}
g_E^{-2}\!=\!({1/3})/({1-\langle P\rangle_{MC}/3}),
\end{equation}
with $\langle P\rangle_{MC}$ the Monte Carlo determined zero temperature plaquette at the critical coupling. The constant in equation (\ref{eq:scaling}) is fixed by the data at $N_t\!=\!12$. 

The leading-order lattice-distorted two-loop scaling ($g_E$-$2l$LD) \cite{Allton:1996kr}, which corrects for lattice artefacts due to the finiteness of $a$, 
\be
\label{eq:scaling_LD}
R_{LD}^{-1}(g_E) = R^{-1}(g_E)\left [1-h\frac{R^2(g_E)}{R^2(g_E^c(N_t=12))}\right ]
\ee
is in good agreement with the data for $h\!\ll\!1$. A larger $h$ is needed when the tadpole improved \cite{Lepage:1992xa} (lattice bare) coupling is used in equation (\ref{eq:scaling_LD}), $h\!=\!0.024$ ($h\!=\!0.0275$) and $h\!=\!0.018$ ($h\!=\!0.022$) for $N_f\!=\!6$ and $N_f\!=\!4$, respectively. The increase of $h$ from $N_f\!=\!4$  to $N_f\!=\!6$ is consistent with the fact that the latter theory has a lower critical temperature, hence a larger critical lattice spacing and larger lattice artefacts, for the same $N_t$. However, if contributions to asymptotic scaling beyond two loops are significant (more likely for $N_f\!=\!6$ than $N_f\!=\!4$) the $h$ parameter in the two-loop scaling will be affected by higher loop effects. A complete four-loop analysis in the $\overline{MS}$ scheme is desirable, but we currently lack the full conversion of the lattice bare coupling to the $\overline{MS}$ coupling for the given lattice action. In using the asymptotic scaling formula valid in the massless limit, we have assumed that the relative shift of the (pseudo) critical couplings due to the nonzero fermion mass with varying $N_t$ is within present uncertainties. A more refined study is certainly worthwhile, following the strategy used for the precise determination of the QCD pseudocritical temperature \cite{Aoki:2006we}. 

\section{\label{sec:discussion} Discussion }
This study consistently suggests that the conformal window of QCD opens between $N_f\!=\!6$ and $N_f\!=\!8$ fundamental flavours. The devised strategy makes use of the underlying symmetries of the $SU(3)$ theory and their breaking pattern with varying $N_f$ and temperature, and it should be preliminary to the lattice investigation of other properties.

Consider the hadron spectrum; it is numerically challenging to distinguish between the spectrum just above and just below the lower edge, in the presence of a nonzero fermion mass and at finite volume. For example, the features of the $N_f\!=\!8$ spectrum studied in \cite{Appelquist:2016viq} can be explained by an underlying exact chiral symmetry, in agreement with this work: the lowest-lying scalar meson is light because it needs to be degenerate with its chiral partner, the pseudoscalar, in the chiral limit. One should also observe that inside the conformal window residual finite volume effects behave differently from QCD, given that only a long-range spin-dependent Coulomb potential is present. In fact, the study of $N_f\!=\!12$ \cite{Lombardo:2014pda} has shown that spin-1 channels can be more affected than spin-0 channels by the finite size of the box, leading to an enhancement of the $m_\rho/m_\pi$ finite volume ratio not dissimilar to $N_f\!=\!8$ in \cite{Appelquist:2016viq}.

A lower edge of the QCD conformal window around $N_f\!=\!7$ is not far from the prediction of two-loop perturbation theory,  it is in remarkable agreement with the large-$N$ result $N_f/N\!=\!5/2$, for $N\!=\!3$ \cite{Bochicchio:2013aha} and with four-loop perturbation theory \cite{PhysRevD.83.056011}. It is also consistent with the perturbatively small value of the fermion mass anomalous dimension of the $N_f\!=\!12$ theory \cite{Lombardo:2014pda, Cheng:2013xha}, because $N_f\!=\!12$ is not close to the lower edge of the conformal window. A more accurate study of $N_f\!=\!7$ at a smaller fermion mass should allow to determine if this theory is conformal or confining, while any form of preconformal behaviour, if it exists, is likely to be manifest in the $N_f\!=\!6$ theory. The latter is not far from real world QCD and it remains an instructive playground for standard model extensions, including dark matter. 
\section*{Acknowledgments}
We thank M. Bochicchio, M.P. Lombardo and K. Miura for interesting discussions. The numerical work was in part based on the MILC public lattice gauge theory code and carried out on the Dutch national supercomputer Cartesius with the support of SURF Foundation.

\balance

\bibliography{references.bib}

\begin{thebibliography}{10}
\expandafter\ifx\csname url\endcsname\relax
  \def\url#1{\texttt{#1}}\fi
\expandafter\ifx\csname urlprefix\endcsname\relax\def\urlprefix{URL }\fi
\expandafter\ifx\csname href\endcsname\relax
  \def\href#1#2{#2} \def\path#1{#1}\fi

\bibitem{Caswell:1974gg}
W.~E. Caswell, Phys. Rev. Lett. 33 (1974) 244.

\bibitem{Banks:1981nn}
T.~Banks, A.~Zaks, Nucl. Phys. B196 (1982) 189.

\bibitem{Litim:2014uca}
D.~F. Litim, F.~Sannino, JHEP 12 (2014) 178.

\bibitem{Aharony:1999ti}
O.~Aharony, et~al., Phys. Rept. 323 (2000) 183.

\bibitem{Bochicchio:2016euo}
M.~Bochicchio\href {http://arxiv.org/abs/1606.04546} {\path{arXiv:1606.04546}}.

\bibitem{Bochicchio:2013eda}
M.~Bochicchio, Nucl. Phys. B875 (2013) 621.

\bibitem{Deuzeman:2012ee}
A.~Deuzeman, et~al., Phys. Lett. B720 (2013) 358.

\bibitem{Lombardo:2014pda}
M.~Lombardo, et~al., JHEP 1412 (2014) 183.

\bibitem{daSilva:2012wg}
T.~N. da~Silva, E.~Pallante, PoS LATTICE2012 (2012) 052.

\bibitem{Cheng:2011ic}
A.~Cheng, A.~Hasenfratz, D.~Schaich, Phys. Rev. D85 (2012) 094509.

\bibitem{Deuzeman:2008sc}
A.~Deuzeman, M.~P. Lombardo, E.~Pallante, Phys. Lett. B670 (2008) 41.

\bibitem{Miura:2012zqa}
K.~Miura, M.~P. Lombardo, Nucl. Phys. B871 (2013) 52.

\bibitem{Deuzeman:2009mh}
A.~Deuzeman, M.~P. Lombardo, E.~Pallante, Phys. Rev. D82 (2010) 074503.

\bibitem{Aoki:2012eq}
Y.~Aoki, et~al., Phys. Rev. D86 (2012) 054506.

\bibitem{Cheng:2013xha}
A.~Cheng, et~al., Phys. Rev. D90 (2014) 014509.

\bibitem{Deuzeman:2011pa}
A.~Deuzeman, et~al., PoS LATTICE2011 (2011) 321.

\bibitem{Parisi1981}
G.~Parisi, AIP Conf. Proc. No. 68 (1981) 1531.

\bibitem{Allton:1996kr}
C.~R. Allton\href {http://arxiv.org/abs/hep-lat/9610016}
  {\path{arXiv:hep-lat/9610016}}.

\bibitem{Lepage:1992xa}
G.~P. Lepage, P.~B. Mackenzie, Phys. Rev. D48 (1993) 2250.

\bibitem{Aoki:2006we}
Y.~Aoki, et~al., Nature 443 (2006) 675.

\bibitem{Appelquist:2016viq}
T.~Appelquist, et~al., Phys. Rev. D93 (2016) 114514.

\bibitem{Bochicchio:2013aha}
M.~Bochicchio\href {http://arxiv.org/abs/1312.1350} {\path{arXiv:1312.1350}}.

\bibitem{PhysRevD.83.056011}
T.~A. Ryttov, R.~Shrock, Phys. Rev. D83 (2011) 056011.

\end{thebibliography}

\end{document}